\documentclass[showpacs,preprintnumbers,twocolumn,
amsmath,amssymb,groupedaddress,superscriptaddress]{revtex4}
\usepackage{graphicx}
\usepackage{dcolumn}
\usepackage{bm}

\begin{document}


\title{Superscaling and Charge-Changing Neutrino Scattering from
Nuclei in\\ the $\boldsymbol \Delta$-Region beyond the Relativistic
Fermi Gas Model}

\author{M.V. Ivanov}
\affiliation{Institute for Nuclear Research and Nuclear Energy,
Bulgarian Academy of Sciences, Sofia 1784, Bulgaria}

\author{M.B. Barbaro}
\affiliation{Dipartimento di Fisica Teorica, Universit\`a di Torino
and INFN, Sezione di Torino, Via P. Giuria 1, I-10125 Torino, Italy}

\author{J.A. Caballero}
\affiliation{Departamento de F\'\i sica At\'omica, Molecular y
Nuclear, Universidad de Sevilla, Apdo.~1065, 41080 Sevilla, Spain}

\author{A.N. Antonov}
\affiliation{Institute for Nuclear Research and Nuclear Energy,
Bulgarian Academy of Sciences, Sofia 1784, Bulgaria}

\author{E. Moya de Guerra}
\affiliation{Departamento de Fisica At\'omica, Molecular y Nuclear,
Facultad de Ciencias Fisicas, Universidad Complutense de Madrid,
E-28040 Madrid, Spain}

\affiliation{Instituto de Estructura de la Materia, CSIC, Serrano
123, E-28006 Madrid, Spain}

\author{M.K. Gaidarov}
\affiliation{Institute for Nuclear Research and Nuclear Energy,
Bulgarian Academy of Sciences, Sofia 1784, Bulgaria}

\affiliation{Instituto de Estructura de la Materia, CSIC, Serrano
123, E-28006 Madrid, Spain}


\begin{abstract}

The superscaling analysis using the scaling function obtained
within the coherent density fluctuation model is extended to
calculate charge-changing neutrino and antineutrino scattering on
$^{12}$C at energies from 1 to 2~GeV not only in the quasielastic
but also in the delta excitation region. The results are compared
with those obtained using the scaling functions from the
relativistic Fermi gas model and from the superscaling analysis of
inclusive scattering of electrons from nuclei.

\end{abstract}


\maketitle

\section[]{INTRODUCTION\label{sect1ant}}

The analyses of scaling
(\emph{e.g.}~\cite{ant01,ant02,ant03,ant04,ant05,ant06,ant07,ant08,ant09,ant10})
and superscaling
(\emph{e.g.}~\cite{ant10,ant11,ant12,ant13,ant14,ant15,ant16,ant17,ant18,ant19,ant21,ant20})
phenomena observed in electron scattering from nuclei and the
consideration on the same basis of neutrino (antineutrino)-nucleus
scattering are among the important tasks of nuclear physics in the
last decades. Scalings of the first and the second kind (a very weak
dependence of the reduced cross section on the momentum transfer $q$
and on the mass number, respectively) at excitation energies below
the quasielastic (QE) peak turn out to be related to the
high-momentum components of the nucleon momentum distribution $n(k)$
at $k > 2$~fm$^{-1}$ which are similar for all nuclei and are due to
the short-range and tensor correlations in the nuclei. One says that
the reduced cross sections exhibits superscaling when both types of
scaling occur. The violation of the scaling of the first kind above
the QE peak is related to the excitation of a nucleon in the nucleus
to a $\Delta$-resonance~\cite{ant15,ant22} and to effects of the
meson exchange currents~\cite{ant23,ant24,ant25,ant26,ant27}. The
first theoretical explanations of the superscaling have been given
in~\cite{ant10,ant11} in the framework of the relativistic Fermi gas
(RFG) model. The analyses of the world data on inclusive
electron-nucleus scattering in~\cite{ant12,ant13} confirmed the
observation of this phenomenon, but simultaneously they showed the
necessity to consider it on the basis of more complex dynamical
picture of finite nuclear systems beyond the RFG. The main reason
for this is that the scaling function in the RFG model is
$f_\text{RFG}^\text{QE}(\psi ') = 0$ for $\psi ' \leq -1$, whereas
the experimental scaling function extracted from $(e,e')$ data
extends to large negative values of the scaling variable $\psi '$ up
to $\psi ' \approx -2$ where effects beyond the mean-field
approximation are important. A theoretical approach that correctly
interprets superscaling in the $\psi ' \leq 0$ region is the
coherent density fluctuation model (CDFM)
(\emph{e.g.},~\cite{ant28,ant29}). This model represents a natural
extension of the Fermi gas model to realistic nuclear systems and it
is based on the generator coordinate method~\cite{ant30}. In the
CDFM the QE scaling function $f(\psi ')$ is related to realistic
nucleon momentum and density distributions and it agrees with the
data for negative values of $\psi '$, including $\psi '\lesssim -1$
\cite{ant16,ant17,ant18,ant19,ant20}. This is so because the CDFM
momentum distribution is not a sharp function of $k$ as the RFG one
is. Reliable separation of inclusive electron scattering data into
their longitudinal and transverse contributions for $A >4$ nuclei
made it possible to obtain (see, {\em
e.g.}~\cite{ant12,ant13,ant15,ant34}) from the data a ``universal''
phenomenological QE scaling function $f^\text{QE}(\psi ')$. In the
present work we use the fit of~\cite{ant14}, which is based on the
experimental analysis of J.~Jourdan~\cite{Jourdan:1996ut}. A
striking feature of the scaling function $f^\text{QE}(\psi ')$
extracted from the superscaling analysis (SuSA) is its asymmetric
shape with respect to the peak position $\psi '=0$ with a pronounced
tail extended towards positive $\psi '$ values. This is in contrast
to the scaling function in the RFG model that is symmetric with
respect to $\psi ' = 0$. This property of the phenomenological SuSA
scaling function imposed further theoretical considerations. A
detailed investigation of such asymmetry has been presented in
\cite{ant31,ant32,ant33} in the context of the relativistic mean
field (RMF) approach. These studies have proved the crucial role
played by the description of final-state interactions (FSI), through
the RMF, in order to reproduce adequately the asymmetric shape shown
by the data analysis.

The approach of SuSA to the QE electron scattering (at energies
from several hundred MeV to a few GeV) have been extended to
include also processes in which $\Delta$-excitation
dominates~\cite{ant34}. In the CDFM this was done in~\cite{ant19}.

The validity of superscaling in inclusive electron scattering
allowed one to start studies of neutrino (antineutrino) scattering
off nuclei on the same basis (\cite{ant31,ant34,ant35,ant36}).
Given the corresponding scaling functions, the cross sections of
charge-changing (CC)~\cite{ant31} or neutral-current
(NC)~\cite{ant27} neutrino (antineutrino) -- nucleus scattering
cross sections for intermediate to high energies can be obtained
by multiplying the elementary single-nucleon (s.n.) CC or NC
neutrino (antineutrino) cross sections by the corresponding
scaling function. This procedure relies on some assumptions which
have recently been tested within the RMF+FSI model and are related
to the isospin degrees of freedom~\cite{ant37}.

A number of other theoretical studies of CC
(\emph{e.g.}~\cite{ant38,ant39,ant40,ant41,ant42,ant43,ant44,ant45,ant46,ant47})
and NC
(\emph{e.g.}~\cite{ant38,ant39,ant46,ant48,ant49,ant50,ant51})
neutrino (antineutrino)-nucleus scattering has also been developed
in recent years.

In the QE region the CDFM scaling function (with asymmetry
introduced in~\cite{ant19}) has been applied to analyze
charge-changing neutrino (antineutrino) scattering on $^{12}$C
(for energies of the incident particles from 1 to 2~GeV)
in~\cite{ant19} and neutral current neutrino (antineutrino)
scattering on the same nucleus with proton and neutron knockout
in~\cite{ant52}. The results were compared with those from the RFG
model and from the superscaling analysis
(SuSA)~\cite{ant15,ant34}. These analyses made it possible to gain
information simultaneously and on the same footing about the role
of both the local density and the momentum distribution in nuclei
for the description of superscaling and of electron- and
neutrino-nucleus scattering ({\em e.g.}~\cite{ant17,ant19}). One
of the advantages of the superscaling analysis within the CDFM was
to find the relationship~\cite{ant17} between the behavior of the
scaling function for negative values of $\psi'$ and the slope of
the nucleon momentum distribution $n(k)$ at higher values of the
momentum ($k > 1.5$~fm$^{-1}$) which is similar for all nuclei due
to the short-range nucleon-nucleon correlations. It became
possible to show the sensitivity of the calculated CDFM scaling
function to the peculiarities of $n(k)$ in different regions of
the momentum~\cite{ant17}. It was also shown that the existing
data on the $\psi'$-scaling are informative for $n(k)$ at momenta
up to $k \leq 2$--$2.5$~fm$^{-1}$.

The aim of this work is to extend the CDFM scaling approach from the
QE-region to the $\Delta$-region for CC neutrino and antineutrino
scattering from nuclei using a constructed realistic CDFM scaling
function for the same region.

The article is organized in the following way: the theoretical
scheme is given in Section~\ref{sect2ant}. It includes the main
relationships of the CDFM scaling functions both in QE- and
$\Delta$-region as well as a brief outline of the formalism for CC
neutrino scattering. The results for $^{12}$C($\nu_{\mu},\mu^{-}$)
and $^{12}$C($\bar{\nu}_{\mu},\mu^{+}$) reaction cross sections
are presented and discussed in Section~\ref{sect3ant}. The
conclusions are summarized in Section~\ref{sect4ant}.

\section[]{THE THEORETICAL SCHEME\label{sect2ant}}

\subsection[]{CDFM scaling function in the QE region\label{sect2Aant}}

The QE CDFM scaling function was
obtained~\cite{ant16,ant17,ant18,ant19} on the basis of the local
density distribution, $\rho(r)$, as well as on the basis of the
nucleon momentum distribution, $n(k)$. It is expressed by the sum
of the proton $f_{p}^\text{QE}(\psi^{\prime})$ and neutron
$f_{n}^\text{QE}(\psi^{\prime})$ scaling functions, which are
determined by the proton and neutron densities, $\rho_{p}(r)$ and
$\rho_{n}(r)$, (or by corresponding momentum distributions
$n_{p}(k)$ and $n_{n}(k)$), respectively~\cite{ant19}:
\begin{equation}
f^{QE}(\psi^{\prime})=\dfrac{1}{A}[Zf_p^\text{QE}(\psi^{\prime})+Nf_n^\text{QE}(\psi^{\prime})].\label{eq01ant}
\end{equation}
The proton and neutron scaling functions in Eq.~(\ref{eq01ant}) are
presented as sums of scaling functions for negative
[$f_{p(n),1}^\text{QE}(\psi^{\prime})$] and positive
[$f_{p(n),2}^\text{QE}(\psi^{\prime})$] values of $\psi^\prime$:
\begin{equation}
f_{p(n)}^\text{QE}(\psi^{\prime})=
f_{p(n),1}^\text{QE}(\psi^{\prime})+f_{p(n),2}^\text{QE}(\psi^{\prime}),
\label{eq02ant}
\end{equation}
where (in the case when the scaling function is obtained on the
basis of the density distributions)
\begin{multline}
f_{p(n),1}^\text{QE}(\psi^{\prime})=\!\!\!\!\int\limits_{0}^{\alpha_{p(n)}/(k^{p(n)}_{F}
|\psi^{\prime}|)}\!\!\!\!dR
|F_{p(n)}(R)|^{2}f_\text{RFG,1}^{p(n)}(\psi'(R)),\\
\psi^\prime\leq0, \label{eq03ant}
\end{multline}
\begin{multline}
f_{p(n),2}^\text{QE}(\psi^{\prime})=\!\!\!\!
\int\limits_{0}^{c_{2}\alpha_{p(n)}/(k_{F}^{p(n)}\psi^{\prime})}\!\!\!\!\!\!
dR
|F_{p(n)}(R)|^{2} f_\text{RFG,2}^{p(n)}(\psi'(R)),\\
\psi^{\prime}\geq 0, \label{eq04ant}
\end{multline}
with
\begin{equation}
f_\text{RFG,1}^{p(n)}(\psi^\prime(R)) = c_1\left[ 1-\left(
\frac{k^{p(n)}_FR|\psi^\prime|}{\alpha_{p(n)}}
\right)^{2}\right],~\psi^\prime\leq0 \label{eq05ant}
\end{equation}
and with two forms of $f_\text{RFG,2}^{p(n)}(\psi^\prime(R))$: a
parabolic form,
\begin{equation}
f_\text{RFG,2}^{p(n)}(\psi^\prime(R)) =
c_{1}\left[1-\left(\frac{k_{F}^{p(n)}R\psi^{\prime}}{c_2\alpha_{p(n)}}\right
)^{2}\right ],~\psi^\prime\geq0 \label{eq06ant}
\end{equation}
and an exponential form,
\begin{equation}
f_\text{RFG,2}^{p(n)}(\psi^\prime(R)) = c_1\exp\left[
-\frac{k_{F}^{p(n)}R\psi^\prime}{c_2\alpha_{p(n)}}
\right],~\psi^\prime\geq0. \label{eq07ant}
\end{equation}

The normalizations of the functions are:
\begin{gather}
\int\limits_{0}^{\infty}|F_{p(n)}(R)|^{2}dR=1, \label{eq12ant}\\
\int\limits_{-\infty}^{\infty}f_{p(n)}^\text{QE}(\psi^{\prime})d\psi^{\prime}=1,
\label{eq13ant}\\
\int\limits_{-\infty}^{\infty}f^\text{QE}(\psi^{\prime})d\psi^{\prime}=1.\label{eq14ant}
\end{gather}
It can be seen that due to the normalization
conditions~(\ref{eq13ant}) and~(\ref{eq14ant}) the two parameters
$c_1$ and $c_2$ are not independent. In the case of the parabolic
form of $f_\text{RFG,2}^{p(n)}$ [Eq.~(\ref{eq06ant})]
$c_2=\dfrac{3}{2c_1}-1$ and in the case of the exponential form
[Eq.~(\ref{eq07ant})] $c_2= \dfrac{1-({2}/{3})c_1}{0.632c_1}$.

In Eqs.~(\ref{eq03ant}) and (\ref{eq04ant}) the proton and neutron
weight functions are obtained from the proton and neutron
densities, respectively:
\begin{gather}
\left|F_{p(n)}(R)\right|^2=-\dfrac{4\pi
R^3}{3Z(N)}\left.\dfrac{d\rho_{p(n)}(r)}{dr}\right|_{r=R},
\label{eq08ant}\\
\alpha_{p(n)}=\left[\dfrac{9\pi Z(N)}{4}\right]^{1/3},
\label{eq09ant}
\end{gather}
with normalization
\begin{equation}
\int\limits_{0}^{\infty}\rho_{p(n)}(\mathbf{r})d\mathbf{r}=Z(N).
\label{eq10ant}
\end{equation}
In the CDFM the Fermi momentum for the protons and neutrons can be
calculated using the expression
\begin{equation}
k_{F}^{p(n)}=\alpha_{p(n)}\int\limits_{0}^{\infty}dR
\frac{1}{R}|{F}_{p(n)}(R)|^{2}. \label{eq11ant}
\end{equation}

The QE electron scattering was considered within the CDFM
in~\cite{ant19}. Two types of experimental data were considered. In
the first one the transferred momentum in the position of the
maximum of the QE peak extracted from data
($\omega^\text{QE}_\text{exp}$) is
$q_{exp}^\text{QE}\geq450~\text{MeV/c}\approx2k_F$ and thus
corresponds {\em to the domain where scaling is
fulfilled}~\cite{ant15,ant34}. It was found by fitting to the
maximum of the QE peak extracted from data the value of $c_1$ to be
$0.72$--$0.73$, {\em i.e.}, that it is similar to that in the RFG
model case (case of symmetry of the RFG and of the CDFM QE scaling
functions with $c_1= 0.75$). This leads to an almost symmetric form
of the CDFM scaling function for cases in which
$q_{exp}^\text{QE}\geq450~\text{MeV/c}$. In the second type of
experimental data $q_{exp}^\text{QE}$ {\em is not in the scaling
region} ($q_{exp}^\text{QE}\leq450~\text{MeV/c}\approx2k_F$). For
them it was found by fitting to the maximum of the QE peak the value
of $c_1$ to be $0.63$. For these cases the scaling function in the
CDFM is definitely asymmetric. It was shown in~\cite{ant19} that the
results for the almost symmetric CDFM scaling function
$f^\text{QE}(\psi')$ with $c_1 = 0.72$ agree with the data in the
region of the QE peak in cases when
$q_{exp}^\text{QE}\geq450~\text{MeV/c}\approx2k_F$ and overestimates
them when $q_{exp}^\text{QE}\leq450~\text{MeV/c}$. The results
obtained when an asymmetric scaling function $f^\text{QE}(\psi')$
with $f_\text{RFG,2}^{p(n)}(\psi^\prime(R))$ from
Eq.~(\ref{eq06ant}) and the value $c_1 = 0.63$ are used agree with
the data in cases when
$q^\text{QE}_\text{exp}\leq450~\text{MeV/c}\approx2k_F$ and
underestimate them when $q^\text{QE}_\text{exp}\geq450~\text{MeV/c}$
in the region close to the QE peak. So, we pointed out that the two
different values of $c_1$ ($0.72$ and $0.63$) found by the fitting
to the position of $q^\text{QE}_\text{exp}$ (and the corresponding
to them almost symmetric and definitely asymmetric forms of the CDFM
scaling function $f^\text{QE}(\psi')$) are in relation to that
whether $q^\text{QE}_\text{exp}$ is in the domain of the scaling
($q^\text{QE}_\text{exp}\geq2k_F$) or it is not
($q^\text{QE}_\text{exp}\leq2k_F$). In connection to this
consideration, in~\cite{ant19} we showed that the cross section
results for CC neutrino (antineutrino) scattering on $^{12}$C using
the asymmetric QE CDFM scaling function $f^\text{QE}(\psi')$
($c_1=0.63$) for incident energies from $1$ to $2$~GeV are close to
those of SuSA~\cite{ant15,ant34} and are different from the RFG
model (where $c_1 = 0.75$) results.

\subsection[]{CDFM scaling function in the $\boldsymbol\Delta$-region\label{sect2Bant}}

The CDFM scaling analysis was extended in~\cite{ant19} to the
$\Delta $-peak region. The CDFM scaling function was written in the
form:
\begin{equation}\label{eq15ant}
f^{\Delta}(\psi'_{\Delta})=\int\limits_{0}^{\infty}dR|F_\Delta(R)|^2f^{\Delta}_{RFG}(\psi'_{\Delta}(R)),
\end{equation}
where the RFG scaling function in the $\Delta $-domain is given
by~\cite{ant34}:
\begin{equation}\label{eq16ant}
f^{\Delta}_{RFG}(\psi'_{\Delta})=\dfrac{3}{4}(1-{\psi'_{\Delta}}^2)\theta(1-{\psi'_{\Delta}}^2)
\end{equation}
and the weight function $|F_\Delta(R)|^2$ is related to the
density distribution:
\begin{equation}
\left|F_{\Delta}(R)\right|^2=-\dfrac{4\pi
R^3}{3A}\left.\dfrac{d\rho(r)}{dr}\right|_{r=R}. \label{new17}
\end{equation}
In Eqs.~(\ref{eq15ant}) and {(\ref{eq16ant}) the shifted scaling
variable $\psi'_{\Delta}$ is expressed by (see,
\emph{e.g.}~\cite{ant34}):
\begin{equation} \label{eq17ant}
\psi _{\Delta}'\!\equiv\! \left[ \!\frac{1}{\xi _{F}}\!\left(\!
\kappa \sqrt{{{\rho}_{\Delta}'}^{2}\!+\!\dfrac{1}{\tau'}
}\!-\!\lambda' {\rho}_{\Delta}'\!-\!1\!\right)\! \right]
^{1/2}\!\!\!\!\!\times\! \left\{\!
\begin{array}{cc}
+1, & \lambda' \geq {\lambda'}_{\Delta}^{0} \\
-1, & \lambda' \leq {\lambda'}_{\Delta}^{0}
\end{array}
\right.,
\end{equation}
where
\begin{gather}
\xi_F\equiv \sqrt{1+\eta_F^2}-1,\qquad \eta_F \equiv
\dfrac{k_F}{m_N}\label{eq18ant}\\
\lambda'=
\lambda-\dfrac{E_{shift}}{2m_N},\qquad\tau'=\kappa^2-\lambda'^2,\label{eq19ant}\\
\lambda =\dfrac{\omega}{2m_{N}},\qquad\kappa =
\dfrac{q}{2m_{N}},\qquad\tau =\kappa ^{2}-\lambda
^{2},\label{eq20ant}\\ {\lambda'}_{\Delta}^{0}=\lambda
_{\Delta}^{0}-\dfrac{E_{shift}}{2m_N},\quad \lambda
_{\Delta}^{0}=\frac{1}{2}\left[ \sqrt{\mu _{\Delta}^{2}+4\kappa
^{2}}-1\right],\label{eq21ant}\\
 \mu _{\Delta}=m_{\Delta }/m_{N},\label{eq22ant}\\
\rho_{\Delta} =1+\dfrac{\left( \mu
_{\Delta}^{2}-1\right)}{4\tau},\quad {\rho}_{\Delta}'
=1+\dfrac{\left( \mu _{\Delta}^{2}-1\right)}{4\tau'}.\label{eq23ant}
\end{gather}
$q$ and $\omega$ being the transferred momentum and energy, and
$m_\Delta$ and $m_N$ the masses of the $\Delta$-resonance and the
nucleon, respectively.

In Eq~(\ref{eq15ant}):
\begin{equation}\label{eq25ant}
{\psi'_{\Delta}}^2(R)\!=\!\dfrac{\left[\!  \kappa
\sqrt{{{\rho}_{\Delta}'}^{2}\!+\!\dfrac{1}{\tau'} }\!-\!\lambda'
{\rho}_{\Delta}'\!-\!1\!\right]}{\left[\!\sqrt{1\!+\!\dfrac{k_F^2(R)}{m_N^2}}\!-\!1\!\right]}
\!\equiv\! t(R).{\psi_\Delta'}^2,
\end{equation}
where
\begin{equation}\label{eq26ant}
t(R)\equiv\dfrac{\left[\sqrt{1+\dfrac{k_F^2}{m_N^2}}-1\right]}
{\left[\sqrt{1+\dfrac{k_F^2(R)}{m_N^2}}-1\right]}, \quad
k_F(R)=\dfrac{\alpha}{R},
\end{equation}
and
\begin{equation}\label{moe}
\alpha=\left(\dfrac{9\pi A}{8}\right)^{1/3}.
\end{equation}

In the CDFM $k_F$ can be calculated using the density
distribution:
\begin{equation}
k_{F}=\alpha\int\limits_{0}^{\infty}dR
\frac{1}{R}|{F}_{\Delta}(R)|^{2},\quad \label{new28}
\end{equation}
where $|{F}_{\Delta}(R)|$ is given by Eq.~(\ref{new17}) and $\alpha$
by Eq.~(\ref{moe}). In an equivalent formulation of the CDFM,
proposed in~\cite{ant17}, the scaling function and the Fermi
momentum can be obtained using the nucleon momentum distribution.

It was shown in~\cite{ant19} that though the functional forms of
$f^{\Delta}(\psi'_\Delta)$ [Eq.~(\ref{eq15ant})], the weight
function $|F_\Delta(R)|^2$ [Eq.~(\ref{new17})] and of $k_F$
[Eq.~(\ref{new28})]  are like in the QE region (see
Eqs.~(\ref{eq03ant}), (\ref{eq04ant}), Eq.~(\ref{eq08ant}),
Eq.~(\ref{eq11ant}), respectively), it cannot be expected that the
parameters of the densities when a $\Delta$-resonance is excited
({\em e.g.} the half-radius $R_\Delta$ and the diffuseness
$b_\Delta$ when Fermi-type distributions have been used) will be
equal to the values of $R$ and $b$ in the QE case. Indeed the
scaling data of the delta peak extracted from the high-quality world
data for inclusive electron scattering (given in~\cite{ant34}) can
be fitted by using for $^{12}$C the effective values
$R_\Delta=1.565$~fm and $b_\Delta=0.420$~fm and a coefficient in the
right-hand side of Eq.~(\ref{eq16ant}) for the RFG scaling function
$f^{\Delta}_\text{RFG}(\psi_{\Delta}')$ equal to $0.54$ instead of
$3/4$. The value of the Fermi momentum $k_F=1.20$~fm$^{-1}$ ensures
the normalization to unity of the function
$f^{\Delta}_\text{RFG}(\psi_{\Delta}')$. As can be seen, the value
of $R_\Delta$ is smaller than that in the description of the QE
superscaling function for $^{12}$C~\cite{ant16,ant17}
($R=2.470$~fm), whereas the value of $b_\Delta$ is the same as $b$
in the QE case.

\subsection[]{Scaling functions and charge-changing neutrino-nucleus reaction cross section\label{sect2Cant}}

Here we present applications of the CDFM QE- and $\Delta$-scaling
function to the calculations of CC neutrino-nucleus reaction cross
sections. We follow the formalism given in~\cite{ant34}. The CC
neutrino cross section in the target laboratory frame is given in
the form
\begin{equation}
\left[ \frac{d^2 \sigma }{d\Omega dk' }\right] _{\chi }\equiv \sigma
_{0} {\cal F}_{\chi }^{2},  \label{eq32ant}
\end{equation}
where $\chi =+$ for neutrino-induced reactions (for example, $\nu_l+
n\rightarrow \ell^{-}+p$, where $\ell =e,\mu ,\tau $) and $\chi =-$
for antineutrino-induced reactions (for example, $\overline{\nu_l
}+p\rightarrow \ell^{+}+n$),
\begin{equation}
\sigma _{0}\equiv \frac{(G \cos \theta_c)^{2}}{2\pi ^{2}} \left[
k^{\prime }\cos \widetilde{\theta }/2 \right]^2 ,  \label{eq33ant}
\end{equation}
where $G=1.16639\times 10^{-5}$~GeV$^{-2}$ is the Fermi constant
and $\theta_c$ is the Cabibbo angle ($\cos \theta_c=$ 0.9741),
\begin{gather}
\tan ^{2}\widetilde{\theta }/2\equiv \frac{|Q^{2}|}{v_{0}},
\label{eq34ant}\\
v_{0}\equiv (\epsilon +\epsilon ^{\prime })^{2}-q^{2}=4\epsilon
\epsilon ^{\prime }-|Q^{2}|.  \label{eq35ant}
\end{gather}

The function ${\cal F}_{\chi }^{2}$ depends on the nuclear structure
and can be written as~\cite{ant34}:
\begin{multline}
{\cal F}_{\chi
}^{2}=[\widehat{V}_\text{CC}R_\text{CC}+2\widehat{V}_\text{CL}R_\text{CL}
+\widehat{V}_\text{LL}R_\text{LL}+\widehat{V}_\text{T}R_\text{T}]+\\
+\chi[2\widehat{V}_\text{T$'$}R_\text{T$'$}]\label{new33}
\end{multline}
that is, as a generalized Rosenbluth decomposition having
charge-charge (CC), charge-longitudinal (CL),
longitudinal-longitudinal (LL) and two types of transverse
(T,T$'$) responses ($R$'s) with the corresponding leptonic
kinematical factors ($V$'s) presented in~\cite{ant34}. The nuclear
response functions in both QE- and $\Delta$-regions are expressed
in terms of the nuclear tensor $W^{\mu\nu}$ in the corresponding
region, using its relationships with the RFG model scaling
functions. The basic relationships used to calculate the s.n.
cross sections are given in~\cite{ant34}. This concerns the
leptonic and hadronic tensors and the response and structure
functions. In our calculations of neutrino-nucleus cross sections
(following~\cite{ant34}) we use for the nucleon form factors the
H{\"o}hler parametrization 8.2 \cite{ant53} in the vector sector
and the form factors given in~\cite{ant34} in the axial-vector
sector.

In the present work, instead of the RFG functions in the QE and
$\Delta$ regions, we use those obtained in the CDFM (described in
subsections~\ref{sect2Aant} and~\ref{sect2Bant} of this section).

\section[]{RESULTS OF CALCULATIONS AND DISCUSSION\label{sect3ant}}

In this section we present firstly the QE- and $\Delta $- CDFM
scaling functions $f^\text{QE}(\psi_\text{QE}')$ and
$f^{\Delta}(\psi_{\Delta}')$ by means of which the cross sections
of CC neutrino (antineutrino) scattering on $^{12}$C are
calculated.

In this work we have not considered Coulomb distortion of the
outgoing muon. Checks made within the effective momentum
approach~\cite{ant34,ant41,ant54} have shown that these effects are
within a few percent for the high energy muon kinematics and the
light target $^{12}$C considered in this work. Therefore, our
general conclusions about scaling are not modified.

In Fig.~\ref{fig1ant} we compare the QE- and $\Delta $- CDFM
scaling functions, while in Fig.~\ref{fig2ant} a comparison of the
$\Delta $-region CDFM scaling function
$f^{\Delta}(\psi_{\Delta}')$ with the averaged experimental data
for $f^{\Delta}(\psi_{\Delta}')$ taken from~\cite{ant31} (see also
Fig.~5 of~\cite{ant19}) is given.

\begin{figure}[t]
\centering
\includegraphics[width=80mm]{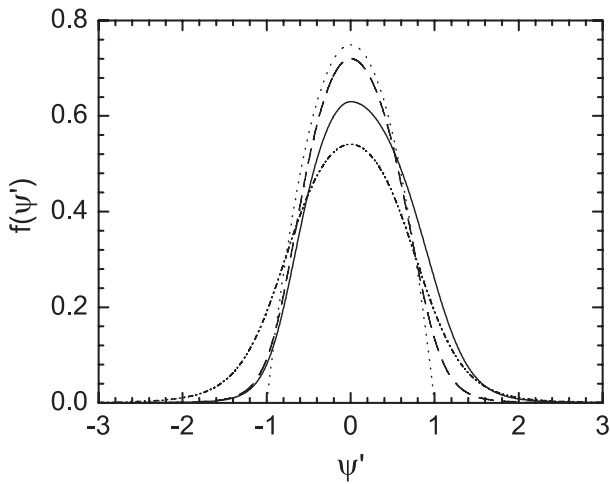}
\caption[]{The CDFM scaling function in the $\Delta$-region
$f^{\Delta}(\psi_{\Delta}')$ (double-dot dashed line) calculated
with $R_\Delta=1.565$~fm, $b_\Delta=0.420$~fm,
$k_F=1.20$~fm$^{-1}$, and a coefficient in the right-hand side of
Eq.~(\ref{eq16ant}) equal to $0.54$ (instead of $3/4$). By dotted,
dashed and solid lines are presented the QE scaling functions
$f^\text{QE}(\psi_\text{QE}')$ in the RFG model and in the CDFM
with $c_1=0.72$ and $c_1=0.63$, respectively.\label{fig1ant}}
\end{figure}

In Figs.~\ref{fig3ant}--\ref{fig7ant} (panels (a)) we give  the
results of calculations for cross sections (the QE- and
$\Delta$-contributions) of neutrino ($\nu_\mu$,$\mu^{-}$)
scattering on $^{12}$C at different muon angles and incident
neutrino energies from 1 to 2~GeV. In the calculations we used the
CDFM scaling function in the QE region
[Eqs.~(\ref{eq01ant})--(\ref{eq06ant}),
(\ref{eq08ant})--(\ref{eq14ant}), using the parabolic form
[Eq.~({\ref{eq06ant}})] of
$f_\text{RFG,2}^{p(n)}(\psi^\prime(R))$] and in the
$\Delta$-region [Eqs.~(\ref{eq15ant})--(\ref{new28})]. The results
of the CDFM in the QE case are compared with those from the RFG
model and SuSA~\cite{ant15,ant34}. We present also (in panels (b))
the sum of the QE- and $\Delta $-contributions to the cross
sections. As an example in Fig.~\ref{fig8ant} ((a) and (b)) we
give the results of the calculations for cross sections of
antineutrino ($\overline{\nu}_\mu$,$\mu^{+}$) scattering on
$^{12}$C for the case of muon angle $\theta_\mu=45^\circ$ and the
incident antineutrino energy $\varepsilon_{\overline{\nu}}=1$~GeV.

\begin{figure}[b]
\centering
\includegraphics[width=80mm]{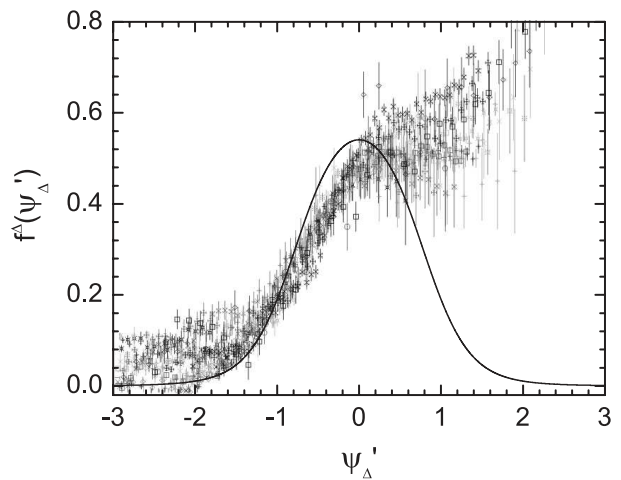}
\caption[]{The same as in Fig.~\ref{fig1ant} for the CDFM  scaling
function $f^{\Delta}(\psi_{\Delta}')$ in the $\Delta$-region
(solid line). Averaged experimental values of
$f^{\Delta}(\psi_{\Delta}')$ are taken
from~\cite{ant34}.\label{fig2ant}}
\end{figure}

\begin{figure*}
\centering
\includegraphics[width=147mm]{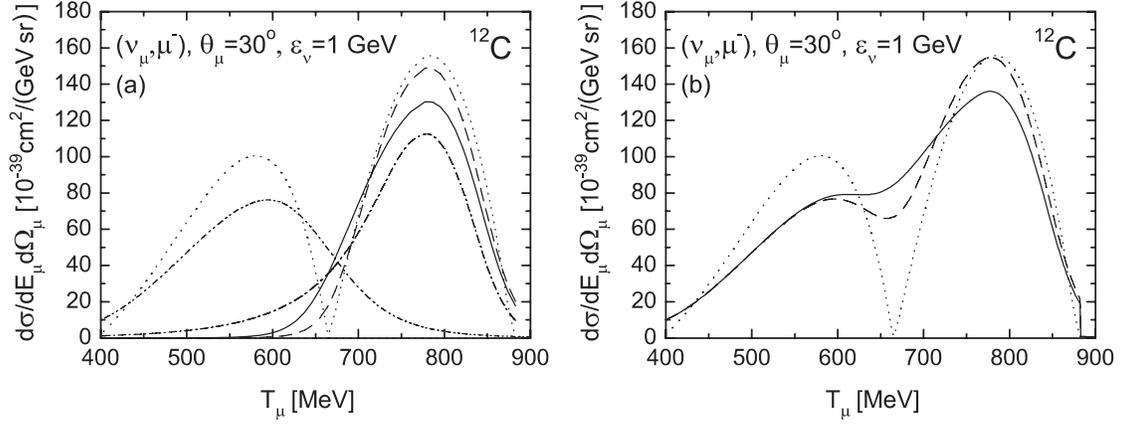}
\caption[]{The cross section of charge-changing neutrino
($\nu_\mu$,$\mu^{-}$) reaction on $^{12}$C at $\theta_\mu=30^\circ$
and $\varepsilon_{{\nu}}=1$~GeV. (a) QE contributions: the result of
CDFM with $c_1=0.63$ (solid line); CDFM with $c_1=0.72$ (dashed
line); RFG (dotted line); SuSA result (dot-dashed line); the result
for the $\Delta $-contribution from the CDFM (double dot-dashed
line). (b) the sum of QE- and $\Delta $-contributions in RFG model
(dotted line), in the CDFM with $c_1=0.63$ (solid line) and
$c_1=0.72$ (dashed line). Here and in the following figures the
range of variation of $\psi^\prime$ and $\psi^\prime_\Delta$ is
approximately $(-2.0,5.5)$ and $(-3.5,+2.5)$,
respectively.\label{fig3ant}}
\end{figure*}

\begin{figure*}
\centering
\includegraphics[width=147mm]{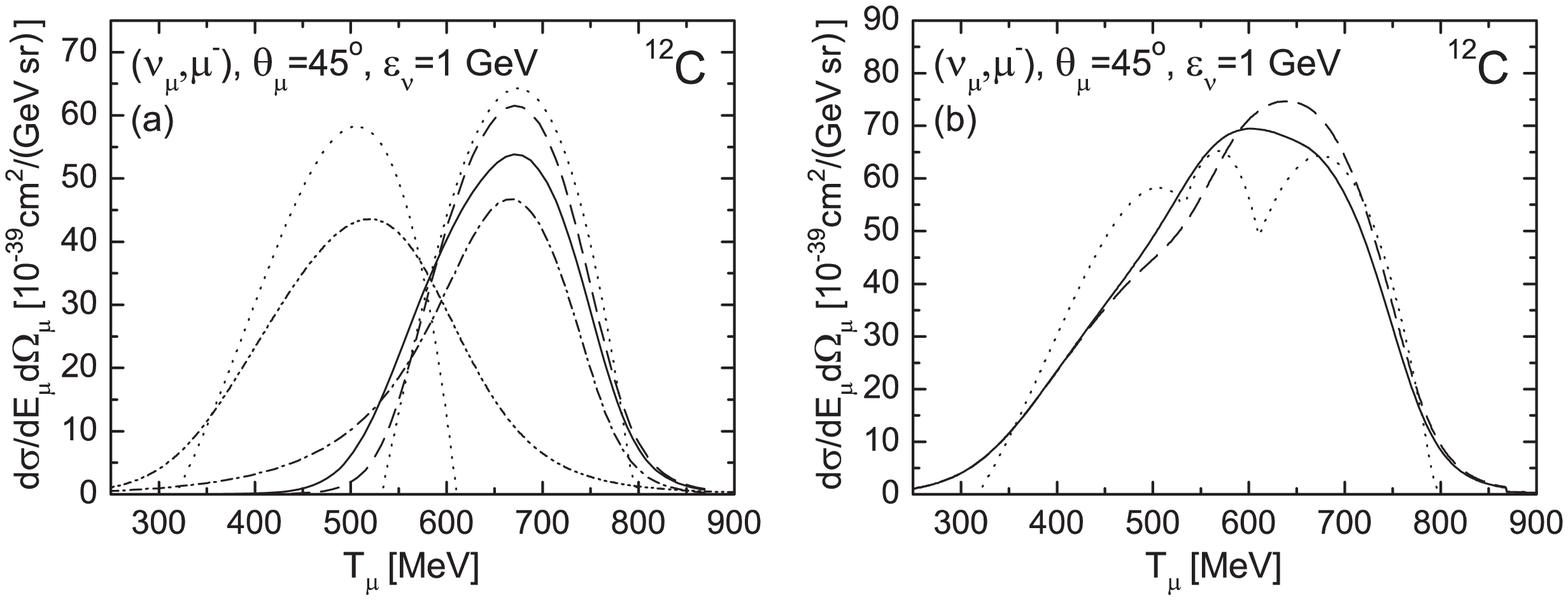}
\caption[]{The same as in Fig.~\ref{fig3ant} for
$\theta_\mu=45^\circ$ and $\varepsilon_{{\nu}}=1$~GeV.
\label{fig4ant}}
\end{figure*}

\begin{figure*}
\centering
\includegraphics[width=150mm]{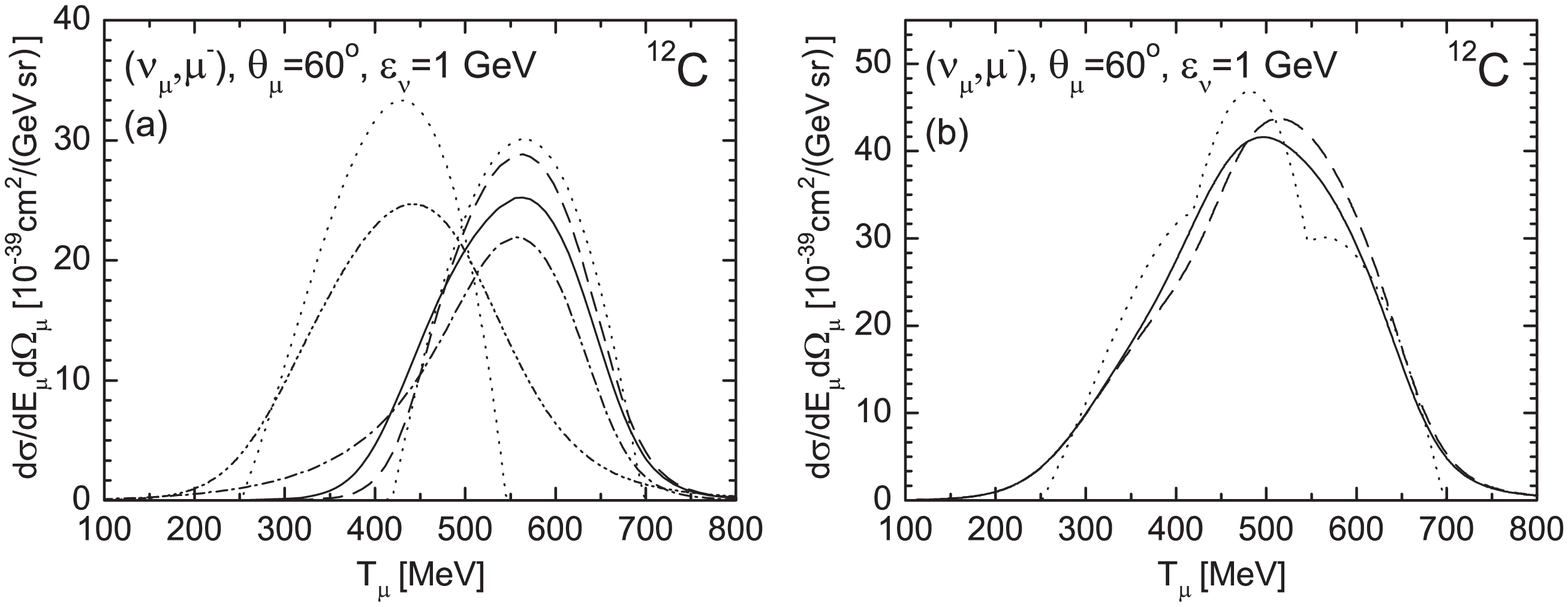}
\caption[]{The same as in Fig.~\ref{fig3ant} for
$\theta_\mu=60^\circ$ and $\varepsilon_{{\nu}}=1$~GeV.
\label{fig5ant}}
\end{figure*}

\begin{figure*}
\centering
\includegraphics[width=150mm]{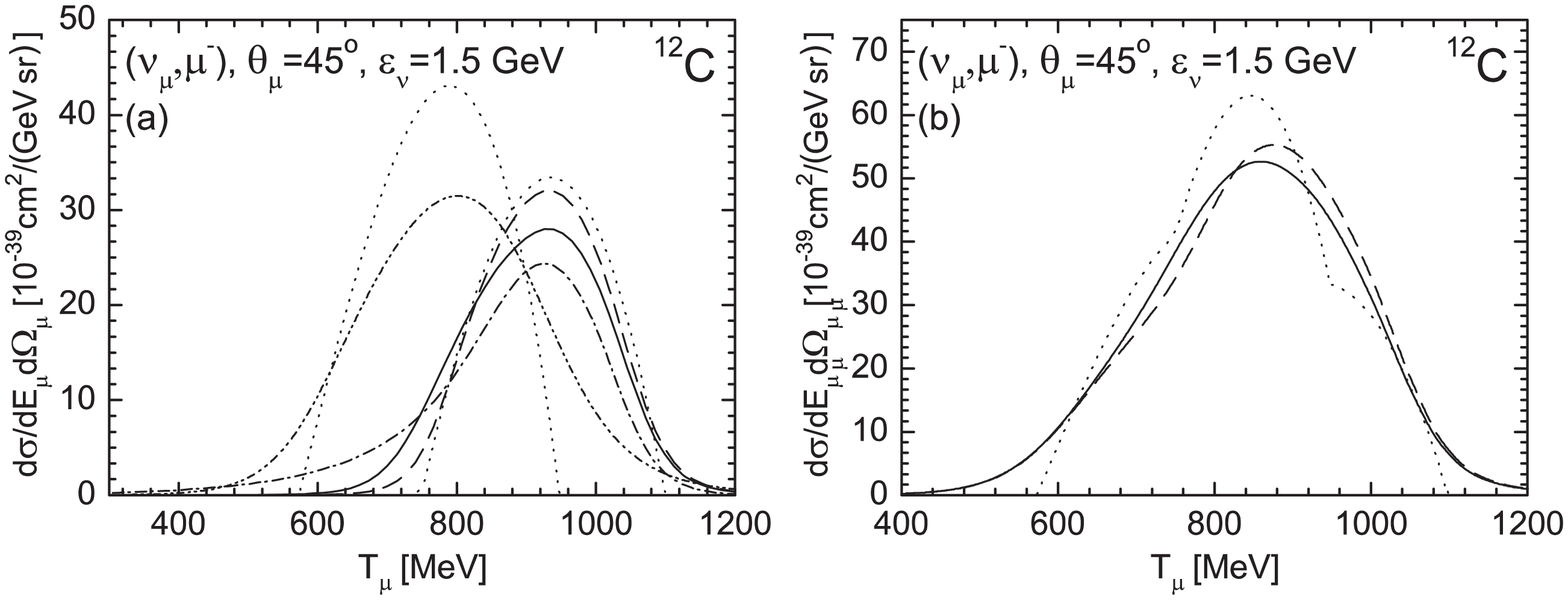}
\caption[]{The same as in Fig.~\ref{fig3ant} for
$\theta_\mu=45^\circ$ and $\varepsilon_{{\nu}}=1.5$~GeV.
\label{fig6ant}}
\end{figure*}

\begin{figure*}
\centering
\includegraphics[width=150mm]{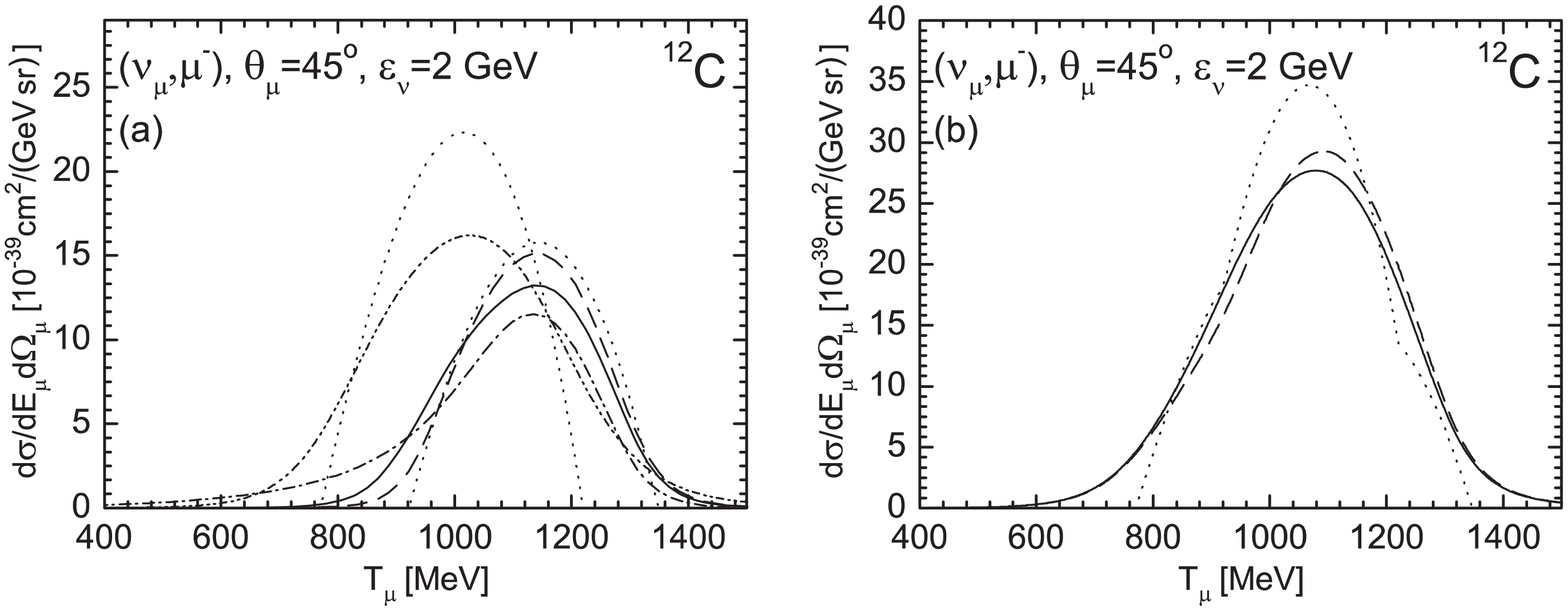}
\caption[]{The same as in Fig.~\ref{fig3ant} for
$\theta_\mu=45^\circ$ and $\varepsilon_{{\nu}}=2$~GeV.
\label{fig7ant}}
\end{figure*}

\begin{figure*}
\centering
\includegraphics[width=150mm]{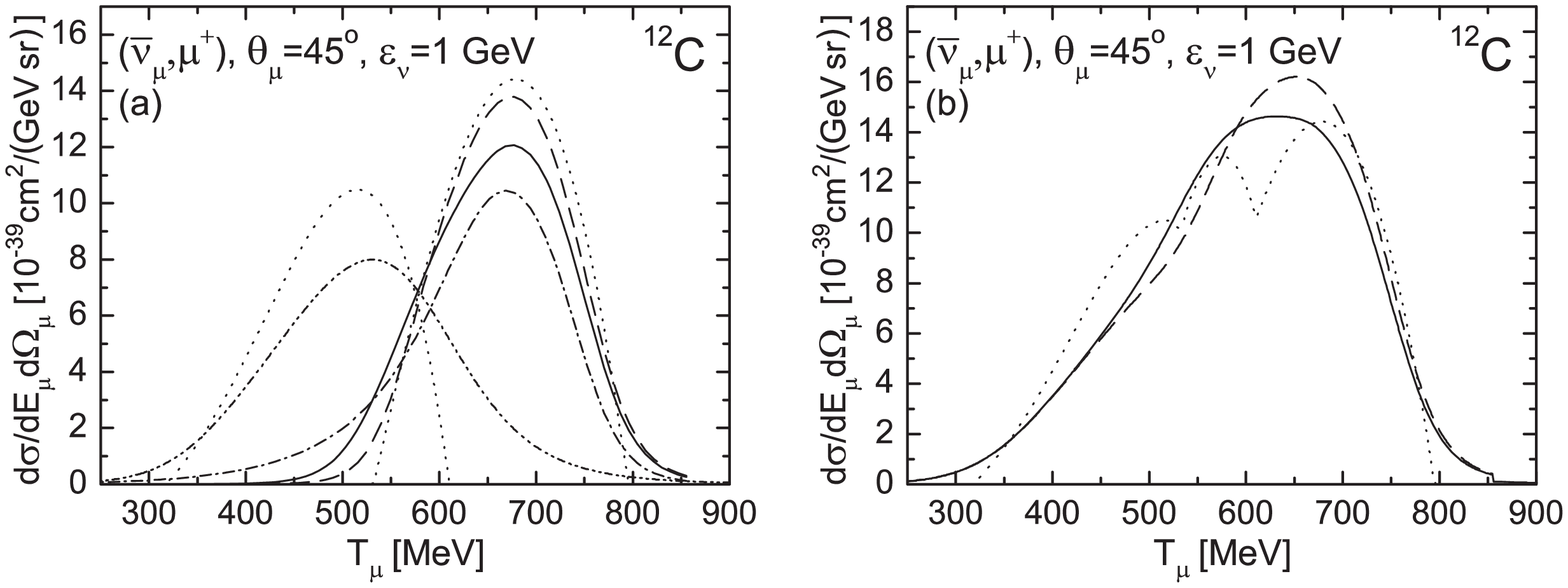}
\caption[]{The cross section of charge-changing antineutrino
($\overline{\nu}_\mu$,$\mu^{+}$) reaction on $^{12}$C at
$\theta_\mu=45^\circ$ and $\varepsilon_{{\nu}}=1$~GeV. The
notations are the same as in Fig.~\ref{fig3ant}. \label{fig8ant}}
\end{figure*}

First, it can be seen from Figs.~\ref{fig3ant}--\ref{fig8ant}
(panels (a)) that the CDFM results (with $c_1=0.72$) for the cross
sections in the QE region are close to those of the RFG model,
while the results of CDFM (with $c_1=0.63$) are between those of
RFG and SuSA. This result could be expected due to the
peculiarities of the QE CDFM scaling function $f(\psi')$, namely,
that when $c_1=0.72$ is used it is similar to that of the RFG
model (see Fig.~\ref{fig1ant}), whereas when $c_1=0.63$ the CDFM
scaling function is closer to that of the SuSA. These properties
of the QE CDFM scaling function were shown in comparison with the
experimental data from the electron scattering and with the RFG
model and SuSA results in Fig.~6 of~\cite{ant52}. This
consideration should be kept in mind also in relation to the
observations from the CDFM analyses of the QE electron scattering
(mentioned above) about the necessity to use almost symmetric
scaling function ($c_1 = 0.72$) or asymmetric one ($c_1 = 0.63$)
at different kinematical conditions. Second, we find that, in
general, the strength of the QE peak decreases with increasing
outgoing angle ($\theta_\mu$) or with increasing incoming energy
($\varepsilon_{{\nu}}$). Third, the height of the $\Delta$-peak
also decreases with increasing $\theta_{\mu}$, but its decrease is
much slower so that, at $\varepsilon_{\nu}=1$ GeV, the relative
height of the two peaks ($\Delta$/QE) goes from about $0.5$ at
$\theta_{\mu}=30^\circ$ to $\geq 1$ at $\theta_{\mu}=60^\circ$.
Fourth, something similar happens when we fix the angle and
increase the energy. The $\Delta$-peak decreases more slowly than
the QE peak. For instance, at $\theta_{\mu}=45^\circ$ the relative
height of the two peaks goes from $\Delta$/QE $\leq 1$ at
$\varepsilon_{\nu}=1$ GeV to $\Delta$/QE $\sim 4/3$ at
$\varepsilon_{\nu}=2$ GeV. Fifth, the overlap between both peaks
is larger with increasing incoming energy and/or increasing
scattering angle. From panels (b) it can be seen for the sum of
the QE- and $\Delta $-contributions that at fixed $\theta_\mu$ the
maximum decreases with the increase of the energy. For
$\theta_\mu=45^\circ$ both CDFM curves (with $c_1=0.63 $ and
$c_1=0.72$) are quite similar for the energies
$\varepsilon_{{\nu}}=1\div2$~GeV. At energy
$\varepsilon_{{\nu}}=1$~GeV and small angles (\emph{e.g.}
$\theta_\mu=30^\circ$) there are two maxima of the cross section,
while at larger angles ($\theta_\mu=45^\circ$ and
$\theta_\mu=60^\circ$) the two peaks merge into one (\emph{e.g.}
for $\varepsilon_{{\nu}}=1$~GeV).

We note also that, as can be seen from Figs.~\ref{fig4ant}
and~\ref{fig8ant}, similarly to the results from~\cite{ant34}, the
antineutrino scattering cross section (for incident energy 1~GeV
and $\theta_\mu=45^\circ$) is about 5 times smaller than the
neutrino one.

Although this is true already at the level of the RFG, the overlap
region is more extended in the present model due to the tails of the
corresponding scaling functions outside the RFG region
$|\psi^\prime|<1$.

\section[]{CONCLUSIONS\label{sect4ant}}

In our work~\cite{ant19} we extended the CDFM superscaling
analysis~\cite{ant16,ant17,ant18} from the QE-region to the
$\Delta$-region of the inclusive electron scattering.
In~\cite{ant19} the CDFM was applied also to charge-changing
neutrino and antineutrino reactions at energies between 1 and
2~GeV from $^{12}$C nucleus \emph{in the quasielastic region}.
Later, in our work~\cite{ant52} we considered neutral current
neutrino and antineutrino scattering with energies of 1~GeV from
$^{12}$C with a proton and neutron knockout using CDFM scaling
functions.

In the present work we use the $\Delta $-scaling functions
obtained within the CDFM in~\cite{ant19} to calculate
charge-changing neutrino and antineutrino scattering \emph{in the
$\Delta$-region} extending our previous QE analysis. So, in this
work we obtain both contributions (in QE- and $\Delta$-region) of
the charge-changing neutrino scattering thus completing the CDFM
analyses of both inclusive electron and neutrino scattering from
nuclei on the same basis, \emph{i.e.} using the same CDFM QE- and
$\Delta $-region scaling functions in both cases, for incident
electrons or neutrino (antineutrino). We consider the scattering
of neutrino (antineutrino) with incident energies between 1 and
2~GeV from the $^{12}$C nucleus at different muon angles. Our
results are compared with those from the RFG model and from
SuSA~\cite{ant15,ant34}. Concerning the QE-contribution to the
cross section we note that the use of asymmetric CDFM scaling
function ($c_1=0.63$) gives results which are close to those from
SuSA, while the symmetric scaling function ($c_1=0.72$) leads to
results similar with the RFG model ones.

The results for the cross sections show the following features: i)
at fixed incident energies the values of the QE- and $\Delta$-peak
maxima decrease with the increase of the muon angle $\theta_\mu$
and the value of the $\Delta$-contribution maximum becomes closer
to that of the QE contribution, ii) at fixed angle $\theta_\mu$
the QE- and $\Delta$-contributions overlap more strongly with the
increase of the neutrino energy and the maximum of the
$\Delta$-peak increases, iii) at fixed angle $\theta_\mu$ the
maximum of the sum of both QE- and $\Delta$-contributions to the
cross section decrease with the increase of the energy. For
$\theta_\mu=45^\circ $ both CDFM curves (with $c_1=0.63$ and
$c_1=0.72$) are quite similar for the interval of neutrino
energies $\varepsilon_\nu=1\div2$~GeV, iv) at energy
$\varepsilon_\nu=1$~GeV and smaller angles (\emph{e.g.}
$\theta_\mu=30^\circ$) there are two maxima of the total sum of
the QE- and $\Delta$-contributions, while at larger angles
($\theta_\mu=45^\circ$ and $60^\circ$) the two peaks merge into
one (for the energy interval $\varepsilon_\nu=1\div2$~GeV), v)
similarly to the results from~\cite{ant34} the antineutrino cross
section (on the example for incident energy 1~GeV and muon angle
of 45 degrees) is about 5 times smaller than the neutrino one.

In summary, it is pointed out that the constructed QE- and
$\Delta$-region scaling functions in the CDFM can be used in a
reliable way for the description of the electron and neutrino
(antineutrino) scattering from nuclei.\vspace*{.3cm}

\begin{acknowledgments}
This work was partly supported by the Bulgarian National Science
Fund under Contracts No.~$\Phi$--1416 and $\Phi$--1501, and by
Ministerio de Educaci\'on y Ciencia (Spain) under contracts
Nos.~FPA2006--13807--C02--01, FIS2005--01105, and FIS2005--00640.
This work is also partially supported by the EU program ILIAS N6
ENTApP WP1. One of the authors (M.K.G.) is grateful for the warm
hospitality given by the CSIC and for support during his stay
there from the State Secretariat of Education and Universities of
Spain (N/Ref.SAB2005--0012).
\end{acknowledgments}


\begin{thebibliography}{99}

\bibitem{ant01}
G. B. West, Phys. Rep. {\bf 18}, 263 (1975).

\bibitem{ant02} I. Sick, D. B. Day, and J. S. McCarthy, Phys.
Rev. Lett. {\bf 45}, 871 (1980).

\bibitem{ant03}
C. Ciofi degli Atti, E. Pace, and G. Salm\`{e}, Phys. Rev. C {\bf
36}, 1208 (1987).

\bibitem{ant04} D. B. Day, J. S. McCarthy, T. W.
Donnelly, and I. Sick, Annu. Rev. Nucl. Part. Sci. {\bf 40}, 357
(1990).

\bibitem{ant05}
C. Ciofi degli Atti, E. Pace, and G. Salm\`{e}, Phys. Rev. C {\bf
43}, 1155 (1991).

\bibitem{ant06}
C. Ciofi degli Atti and S. Simula, Phys. Rev. C {\bf 53}, 1689
(1996).

\bibitem{ant07} C. Ciofi degli Atti and G. B. West, nucl-th/9702009.

\bibitem{ant08} C. Ciofi degli Atti and G. B. West, Phys. Lett. B \textbf{458}, 447 (1999).

\bibitem{ant09} D. Faralli, C. Ciofi degli Atti, and G. B. West, in
\emph{Proceedings of 2nd International Conference on Perspectives in
Hadronic Physics}, ICTP, Trieste, Italy, 1999, edited by S. Boffi,
C. Ciofi degli Atti, and M. M. Giannini (World Scientific,
Singapore, 2000), p.~75.

\bibitem{ant10} W. M. Alberico, A. Molinari, T. W. Donnelly, E. L.
Kronenberg, and J. W. Van Orden, Phys. Rev. C \textbf{38}, 1801
(1988).

\bibitem{ant11} M. B. Barbaro, R. Cenni, A. De Pace, T. W. Donnelly,
and A. Molinari, Nucl. Phys. A \textbf{643}, 137 (1998).

\bibitem{ant12} T. W. Donnelly and I. Sick, Phys. Rev. Lett.
\textbf{82}, 3212 (1999).

\bibitem{ant13} T. W. Donnelly and I. Sick, Phys. Rev. C \textbf{60},
065502 (1999).

\bibitem{ant14} C. Maieron, T. W. Donnelly, and I. Sick, Phys. Rev. C
\textbf{65}, 025502 (2002).

\bibitem{ant15} M.~B.~Barbaro, J.~A.~Caballero, T.~W.~Donnelly, and C.~Maieron,
Phys. Rev. C {\bf 69}, 035502 (2004).

\bibitem{ant16} A. N. Antonov, M. K. Gaidarov, D. N. Kadrev, M. V.
Ivanov, E. Moya de Guerra, and J. M. Udias, Phys. Rev. C
\textbf{69}, 044321 (2004).

\bibitem{ant17} A. N. Antonov, M. K. Gaidarov, M. V.
Ivanov, D. N. Kadrev, E. Moya de Guerra, P. Sarriguren, and J. M.
Udias, Phys. Rev. C \textbf{71}, 014317 (2005).

\bibitem{ant18} A. N. Antonov, M. V. Ivanov, M. K. Gaidarov, E. Moya de Guerra,
P. Sarriguren, and J. M. Udias, Phys. Rev. C \textbf{73}, 047302
(2006).

\bibitem{ant19} A. N. Antonov, M. V. Ivanov, M. K. Gaidarov, E. Moya de Guerra,
J. A. Caballero, M. B. Barbaro, J. M. Udias, and P. Sarriguren,
Phys. Rev. C \textbf{74}, 054603 (2006).

\bibitem{ant20} A. N. Antonov, M. V. Ivanov, M. K. Gaidarov, and E. Moya de
Guerra, Phys. Rev. C \textbf{75}, 034319 (2007).

\bibitem{ant21} O. Benhar, D. Day, and I. Sick, nucl-ex/0603029.

\bibitem{ant22} L. Alvarez-Ruso, M. B. Barbaro, T. W. Donnelly, and
A. Molinari, Nucl. Phys. A \textbf{724}, 157 (2003).

\bibitem{ant23} J. E. Amaro, M. B. Barbaro, J. A. Caballero, T. W.
Donnelly, and A. Molinari, Nucl. Phys. A \textbf{697}, 388 (2002);
\emph{ibid.} \textbf{723}, 181 (2003); Phys. Rept. {\bf 368}, 317
(2002).

\bibitem{ant24} A. De Pace, M. Nardi, W. M. Alberico, T. W. Donnelly,
and A. Molinari, Nucl. Phys. A \textbf{726}, 303 (2003); {\it ibid.}
\textbf{741}, 249 (2004).

\bibitem{ant25}
J.~E.~Amaro, M.~B.~Barbaro, J.~A.~Caballero, T.~W.~Donnelly, and
A.~Molinari, Nucl. Phys. A {\bf 643}, 349 (1998).

\bibitem{ant26}
J.~E.~Amaro, M.~B.~Barbaro, J.~A.~Caballero, T.~W.~Donnelly, and
A.~Molinari, Nucl. Phys. A {\bf 723}, 181 (2003).

\bibitem{ant27}
J.~E.~Amaro, M.~B.~Barbaro, J.~A.~Caballero, and T.~W.~Donnelly,
Phys. Rev. C {\bf 73}, 035503 (2006).

\bibitem{ant28} A.~N.~Antonov, V.~A.~Nikolaev, and I.~Zh.~Petkov,
Bulg. J. Phys. \textbf{6}, 151 (1979); Z. Phys. A \textbf{297}, 257
(1980); \textit{ibid.} \textbf{304}, 239 (1982); Nuovo Cimento A
\textbf{86}, 23 (1985); Nuovo Cimento A \textbf{102}, 1701 (1989);
A.~N.~Antonov, D.~N.~Kadrev, and P.~E.~Hodgson, Phys. Rev. C
\textbf{50}, 164 (1994).

\bibitem{ant29} A.~N.~Antonov, P.~E.~Hodgson, and I.~Zh.~Petkov,
\textit{Nucleon Momentum and Density Distributions in Nuclei}
(Clarendon Press, Oxford, 1988); \textit{Nucleon Correlations in
Nuclei} (Springer-Verlag, Berlin-Heidelberg-New York, 1993).

\bibitem{ant30} J.~J.~Griffin and J.~A.~Wheeler, Phys. Rev. \textbf{108}, 311 (1957).

\bibitem{ant31} J. A. Caballero, J.~E.~Amaro, M. B. Barbaro,
T. W. Donnelly, C. Maieron, and J. M. Udias, Phys. Rev. Lett.
\textbf{95}, 252502 (2005).

\bibitem{ant32} J.A. Caballero, Phys. Rev. C \textbf{74},
015502 (2006).

\bibitem{ant33}
J.~E.~Amaro, M.~B.~Barbaro, J.~A.~Caballero, T.~W.~Donnelly, and
J.~M.~Udias, Phys. Rev. C {\bf 75}, 034613 (2007).

\bibitem{ant34} J. E. Amaro, M. B. Barbaro, J. A. Caballero, T. W. Donnelly,
A. Molinari, and I. Sick, Phys. Rev. C \textbf{71}, 015501 (2005);
M. B. Barbaro, Nucl. Phys. B, Proc. Suppl. {\bf 159}, 186 (2006);
nucl-th/0602011; M. B. Barbaro, J. E. Amaro, J. A. Caballero, and
T. W. Donnelly, in {\it Nuclear Theory: Proceedings of 25th
International Workshop on Nuclear Theory}, Rila Mountains,
Bulgaria, June 26-July 1, 2006, ed. by S. Dimitrova (DioMira,
2006), p. 73.

\bibitem{Jourdan:1996ut} J.~Jourdan, Nucl. Phys. A {\bf 603}, 117 (1996).

\bibitem{ant35} Y. Fukuda {\it et al.} (The Super-Kamiokande
Collaboration), Phys. Rev. Lett. {\bf 81}, 1562 (1998); M. H. Ahn
{\it et al.} (K2K Collaboration), {\it ibid.} {\bf 90}, 041801
(2003); Q.-R. Ahmad {\it et al.} (SNO Collaboration), {\it ibid.}
{\bf 87}, 071301 (2001); {\bf 89}, 011301 (2002); K. Eguchi {\it et
al.} (KamLAND Collaboration), {\it ibid.} {\bf 90}, 021802 (2003);
C. Athanassopoulos {\it et al.} (LSND Collaboration), {\it ibid.}
{\bf 77}, 3082 (1996); {\bf 81}, 1774 (1998).

\bibitem{ant36}
J.~E.~Amaro, M.~B.~Barbaro, J.~A.~Caballero, T.~W.~Donnelly, and
C.~Maieron, Phys. Rev. C {\bf 71}, 065501 (2005).

\bibitem{ant37} J.~A.~Caballero, J.~E.~Amaro, M. B. Barbaro,
T.~W.~Donnelly, and J. M. Udias, Phys. Lett. B {\bf 653}, 366
(2007).

\bibitem{ant38} M. C. Martinez, P. Lava, N. Jachowicz, J.
Ryckebusch, and J. M. Udias, Phys. Rev. C \textbf{73}, 024607
(2006).

\bibitem{ant39} J. Nieves, M. Valverde, and M. J.
Vicente-Vacas, Nucl. Phys. B, Proc. Suppl. \textbf{155}, 263 (2006);
nucl-th/0510010; Phys. Rev. C \textbf{73}, 025504 (2006).

\bibitem{ant40} M. B. Barbaro, J. E. Amaro, J. A. Caballero,
T. W. Donnelly, and A. Molinari, Nucl. Phys. B, Proc. Suppl.
\textbf{155}, 257 (2006); nucl-th/0509022.

\bibitem{ant41} C. Maieron, M. C. Martinez, J. A. Caballero, and
J. M. Udias, Phys. Rev. C \textbf{68}, 048501 (2003).

\bibitem{ant42} A. Meucci, C. Giusti, and F. D. Pacati,
Nucl. Phys. A \textbf{739}, 277 (2004); Nucl. Phys. A \textbf{773},
250 (2006).

\bibitem{ant43} O. Benhar, Nucl. Phys. B, Proc. Suppl. \textbf{139}, 15 (2005); nucl-th/0408045;
 O. Benhar and N. Farina, Nucl. Phys. B, Proc. Suppl. \textbf{139}, 230 (2005); nucl-th/0407106;
 O. Benhar, N. Farina, H. Nakamura, M. Sakuda, and R. Seki, Nucl. Phys. B, Proc. Suppl. \textbf{155}, 254 (2006);
 hep-ph/0510259; Phys. Rev. D \textbf{72}, 053005 (2005).

\bibitem{ant44} G. Co', Nucl. Phys. B, Proc. Suppl. \textbf{159}, 192 (2006);
nucl-th/0601034; A. Botrugno and G. Co', Nucl. Phys. A \textbf{761},
200 (2005); M. Martini, G. Co', M. Anguiano, and A. M. Lallena,
Phys. Rev. C \textbf{75}, 034604 (2007).

\bibitem{ant45} T. Leitner, L. Alvarez-Ruso, and U. Mosel, Phys. Rev. C \textbf{73}, 065502 (2006).

\bibitem{ant46} B. Szczerbinska, T. Sato, K. Kubodera, and T.-S. H. Lee, nucl-th/0610093.

\bibitem{ant47} O. Buss, T. Leitner, L. Alvarez-Ruso, and U. Mosel, arXiv: 0707.0232 [nucl-th] (2007).

\bibitem{ant48} A. Meucci, C. Giusti, and F. D. Pacati,
Nucl. Phys. A \textbf{744}, 307 (2004).

\bibitem{ant49} W.~M.~Alberico and C.~Maieron, arXiv: hep-ph/0210017 (2002).

\bibitem{ant50} T. Leitner, L. Alvarez-Ruso, and U. Mosel, Phys. Rev. C \textbf{74}, 065502 (2006).

\bibitem{ant51} K.~S.~Kim, B.~G.~Yu, M.~K.~Cheoun, T.~K.~Choi, and M.~T.~Chung, arXiv:0707.1370 [nucl-th] (2007).

\bibitem{ant52} A. N. Antonov, M. V. Ivanov, M. B. Barbaro, J. A. Caballero, E. Moya de
Guerra, and M. K. Gaidarov, Phys. Rev. C \textbf{75}, 064617 (2007).

\bibitem{ant53} G.~H\"{o}hler, E.~Pietarinen, I.~Sabba-Stefanescu,
F.~Bor\-kow\-ski, G.~G.~Simon, V.~H.~Walther, and R.~D.~Wendling,
Nucl. Phys. B {\bf 114}, 505 (1976).

\bibitem{ant54} W.~M. Alberico, M.~B. Barbaro, S.~M. Bilenky, J.~A. Caballero, C.
Giunti, C. Maieron, E. Moya de Guerra, and J.~M. Udias, Nucl.Phys. A
\textbf{623}, 471 (1997).

\end{thebibliography}
\end{document}